\documentclass[seceq]{ptptex}
\usepackage{wrapft}

\begin{document}

\markboth{
W. Gl\"ockle, H. Kamada and J. Golak }{
On a translationally invariant nuclear single particle picture}

\title{
On a translationally invariant nuclear single particle picture}

\author{
Walter \textsc{Gl\"ockle}$^{1,}$\footnote{E-mail: Walter.Gloeckle@tp2.ruhr-uni-bochum.de}
Hiroyuki \textsc{Kamada}$^{2,}$\footnote{E-mail: kamada@mns.kyutech.ac.jp
}

and Jacek \textsc{Golak}$^{3,}$}

\inst{
$^1$Institut f\"ur Theoretische Physik II, Ruhr-Universit\"at Bochum, D-4
4780 Bochum, Germany \\
$^2$Department of Physics, Faculty of Engineering, Kyushu Institute of Technology,
Kitakyushu 804-8550, Japan \\

$^3$M. Smoluchowski Institute of Physics, Jagiellonian University, PL-30059 Krak\'ow, Poland
}




\abst{
If one assumes a translationally invariant motion of the nucleons relative
to the c. m. position in single particle mean fields a correlated single particle picture of the
    nuclear wave function emerges. A single particle product ansatz leads
 for that Hamiltonian to nonlinear equations for the single particle wave
 functions. In contrast to a standard not translationally invariant shell model
 picture those single particle s-, p- etc states are coupled.  The strength of
the  resulting coupling is an open question. The Schr\"odinger equation for that
Hamiltonian can be solved by few- and many -body techniques,
which  will allow to check the validity or non-validity of a single  particle product  ansatz.

     Realistic nuclear wave functions  exhibit repulsive 2-body short range correlations.
Therefore a translationally invariant single  particle picture -- if useful at all --
can only be expected beyond those ranges. Since exact A = 3 and 4 nucleon ground state
wave functions and beyond  based on  modern nuclear forces are available, the translationally
invariant
shell model picture can be optimized by an adjustment to the exact wave function and
its validity or non-validity decided.}





\maketitle

\section{Introduction}

The shell model for the nucleus 
has a long tradition.
However, in its
 standard form expressed in single particle variables it is  plagued by violating
translational invariance. Various methods have been suggested to
 remedy this situation, like for instance the generator coordinate method
 \cite{giraud, hill, griffin, wong}.
Clearly, if the shell model is realistic at all,
 the motion of the individual nucleons in a mean field happens in a translationally
invariant manner, namely as  a function of $ \vec u_i \equiv \vec  x_i - \vec X $, where $ \vec x_i $
  are the individual coordinates of particle $i$ and $ \vec X $ is the c. m.
coordinate.  However, this set of coordinate vectors  $ \vec u_i $
 obeys the  obvious condition
   $\sum_{i=1}^A \vec u_i =0 $,  correlating the motion of all particles.

   It is the aim of the present investigation to work out the consequences of
choosing the coordinates $ \vec u_i$ for a shell model picture.

    In Section II we provide some formal basis for this specific choice of coordinates.
We restrict ourselves in this first investigation to systems of three and four
 nucleons.
     Furthermore, our  most simplistic ansatz for the wave function shifts
 the antisymmetry requirement  to the spin-isospin space, which leads to
a symmetric space part under particle permutations.
     Then the very first  ansatz for the space part is
\begin{eqnarray}
\Phi( \vec u_i) = \prod_{ i=1 }^A R(u_i),
\label{1}
\end{eqnarray}
    with $ A=3 $  or 4. Here only s-wave states are assumed.  We denote
 such a form  a correlated single particle picture.

     The nonlinear equations for 3 and 4 particles
     for the state $ R(u) $ assuming a sum of single particle potentials,
 $ V = \sum_{ i=1}^A V(u_i) $, are presented in Section III.

      In the case of the harmonic oscillator potential the nonlinear equations
can be solved analytically and it is shown  that the ansatz (\ref{1})
is indeed the correct one.

    Obviously the question  arises whether the ansatz (\ref{1}) for the
wave function is at all valid for general single particle mean field potentials.
 To that effect the Hamiltonian,
    including the sum of single particle potentials,  can be expressed in
standard Jacobi variables. This is displayed in Section IV.

     The resulting Sch\"rodinger equations for 3 and 4 particles (in this  case
bosons for the space part)  can be solved exactly in the form of the Faddeev-Yakubovsky
      equations, which will be formulated in that Section. Having the exact
wave function  at ones disposal, one can then investigate how well the above
 shell model ansatz (\ref{1})  is realized,
 whether contributions beyond s-wave are needed, or whether that hope
is not realistic at all. An optimization algorithm relating the shell
model ansatz to the exact wave function is presented in Section V.

The numerical investigations for solving  the  nonlinear equations
  for $R(u)$, for solving the Faddeev-Yakubovsky equations for the shell-model
Hamiltonian  and for the optimal extraction of $ R(u)$ from  the
 exact wave functions is left to  forthcoming investigations.

       The main task however remains. The realistic nuclear wave function
is determined by two- and three-nucleon forces. First estimates in
the case of the $\alpha $-particle
indicate that even small contributions from proper 4N forces
\cite{golak,nogga1} are needed. Based on these forces numerically exact wave
functions are nowadays routinely generated
        for three and four nucleons \cite{nogga2,nogga3,kamada1,gloeck1,nogga4,kamada2,
nogga5}. The question however arises,  how well these wave functions for pair
distances larger than a certain  distance $r_0 $ can be represented in the form of
a correlated shell model ansatz like in (\ref{1}), or whether higher partial waves
and more complicated symmetries with respect to
space-, spin- and  isospin parts of the wave function are required. Clearly for
pair distances smaller than $ r_0 $ short range repulsive features
 are present in the realistic wave functions which can not be represented in
the shell model form. On the other hand it is known that the short pair distance
behavior is essentially universal for light nuclei \cite{gloeck2,gloeck3,gloeck4}
aside from proper normalization, which might allow an overall
description: short range depletion and correlated shell model feature at
larger distances.  The value  $ r_0 $ is expected to be somewhat smaller
than  1~fm. Section VI provides some suggestion on how an optimal extraction of
a correlated single particle picture can be obtained from realistic
 three- and four-nucleon wave functions.   We summarize  in Section  VII.

\section{ The Formal Basis}

The translationally invariant single particle coordinates for $n$
particles  are defined as
     \begin{eqnarray}
     \vec u_i \equiv \vec x_i - \frac{1}{n} \sum_{ j=1}^n \vec x_j =
\frac{ n-1}{n} \vec x_i - \frac{1}{n} \sum_{ j\ne i}^n \vec x_j\label{2}
      \end{eqnarray}

      Since  $ \sum_{i=1}^n \vec u_i =0 $, the mapping from the  n $\vec x_i $
to the n $ \vec u_i $ can not  be inverted and we choose the new variables as the
       first (n-1) $\vec u_j$ together with the c.m. coordinate $ \vec X$
\begin{eqnarray}
\vec X = \frac{1}{n} \sum_{j=1}^n \vec x_j .
\end{eqnarray}
      It is a straightforward exercise to express the kinetic energy in
terms of the new variables
\begin{eqnarray}
T = - \frac{1}{2m} \sum_{ k=1}^n \nabla_{x_k}^2 = - \frac{1} {2m} ( \frac{n-1}{n} \sum_{ i=1}^{n-1} \nabla_{u_i}^2 - \frac{1}{n} \sum_{ i\ne j} \vec \nabla_{u_i} \cdot \vec \nabla_{ u_j} )
- \frac{ 1}{ 2mn} \nabla_X^2\label{4}
\end{eqnarray}

Clearly, the first part refers to the  relative motion,   the second part to
the c. m. motion.
      While the choice of Jacobi coordinates avoids mixed terms in the kinetic
energy, here they are unavoidable.

       Let us now restrict ourselves to three and four particles. If one chooses
 a Slater determinant with equal  space-dependent single particle wave functions,
 the symmetric part of the form (\ref{1})
        factors out and one is left with a totally antisymmetric spin-isospin part.
For a proton-proton-neutron ($ppn$) system this has the form
\begin{eqnarray}
      \chi_3 & = &  | ( t=0 \frac{1}{2}) T= \frac{1}{2} > | ( s=1
 \frac{1}{2}) S= \frac{1}{2} >\cr
      &  - &  | ( t=1 \frac{1}{2}) T= \frac{1}{2} > | ( s=0 \frac{1
}{2} ) S= \frac{1}{2} >,
\label{5}
\end{eqnarray}
       where the two-body spin or isospin state is coupled with the spin or
isospin $\frac{1}{2} $ of the third particle to total spin $ S= \frac
{1}{2}$  or total isospin $ T = \frac{1}{2}$.
       This state together with a symmetric space part is known \cite{blank, friar}
as the principal S-state for realistic $^3$He wave functions and contributes
with about 90\%  to the norm. This result by itself clearly indicates that this
choice of the one Slater determinant can not exhaust the full wave function
 but at least a very large portion of it.

       For the $ppnn$ system the totally antisymmetric spin-isospin part of
the wave function has the form
\begin{eqnarray}
      \chi_4 & = &  ( 1 - P_{23} - P_{24}) | (  \frac{1}{2} \frac{1}{2}
) 0  (  \frac{1}{2} \frac{1}{2}) 0 S=0 >\cr
       & & ( 1 + P_{13} P_{24})  | (  \frac{1}{2} \frac{1}{2}) 1 1 >_{12}
 | (  \frac{1}{2} \frac{1}{2}) 1 - 1 >_{34},
\label{6}
\end{eqnarray}
where  $ P_{ij} $ is a transposition  of particles $i$ and $j$. That state has
total spin $ S=0 $ and total isospin $ T=0 $. Again in relation to the norm of a realistic
        $\alpha$-particle wave function it accounts for about 90\%  \cite{nogga4}.

          Now we provide  some formal properties, whose verification is left to the reader.
The Heisenberg commutation  relations
\begin{eqnarray}
[ w_{ k\alpha}, u_{j\beta}] = \delta_{kj} \frac{1}{i} \delta_{\alpha \beta}\label{7}
  \end{eqnarray}
  are obeyed, where $ w_{k\alpha} \equiv \frac{\partial T} {\partial {\dot u}_{k\alpha}}$
are components of the conjugate momenta.

The relative   orbital angular momentum  has the form
\begin{eqnarray}
        \vec L_{rel}  =  \vec u_1 \times \frac{1}{i} \vec \nabla_{u_1 }
 + \vec u_2 \times \frac{1}{i} \vec \nabla_{u_2 },
\label{8}
\end{eqnarray}
which justifies that standard Clebsch-Gordon coupling in the variables
$ \vec u_1 $ and $ \vec u_2 $ can be used.

Using (\ref{4}) for $n=3$,  the translationally invariant shell
 model  Hamiltonian is given by
\begin{eqnarray}
        H_3 = - \frac{1}{3m} ( \nabla_{u_1}^2 + \nabla_{u_2}^2  - \vec
\nabla_{u_1}\cdot \vec \nabla_{u_2}) + V(u_1) + V(u_2) + V(u_3),
\label{9}
\end{eqnarray}
where $ u_3 = | \vec u_1 + \vec u_2| $. Obviously, a separation of variables is not possible.
However, the symmetry of the  kinetic  energy  under particle exchanges is valid:
\begin{eqnarray}
\nabla_{u_1}^2 +   \nabla_{u_2}^2 - \vec \nabla_{u_1}\cdot \vec \nabla_{u_2} & = &  \nabla_{u_2}^2 +   \nabla_{u_3}^2 - \vec \nabla_{u_2} \cdot \vec \nabla_{u_3}\cr
        &  = &   \nabla_{u_3}^2 +   \nabla_{u_1}^2 - \vec \nabla_{u_3}\cdot \vec \nabla_{u_1}.
\label{10}
\end{eqnarray}

 In the case  of  four particles the translationally invariant shell model Hamiltonian is given as
  \begin{eqnarray}
  H_4 & = &  - \frac{3}{8m} ( \nabla_{u_1}^2 +   \nabla_{u_2}^2 + \nabla_{u_3}^2 
  - \frac{2}{3} (  \vec \nabla_{u_1}\cdot \vec \nabla_{u_2}
   + \vec \nabla_{u_1}\cdot \vec \nabla_{u_3}
   +  \vec \nabla_{u_2}\cdot \vec \nabla_{u_3}))\cr
   &  + &  V(u_1) + V(u_2) + V(u_3) + V(u_4),
\label{11}
  \end{eqnarray}
  with $ u_4 = | \vec u_1 + \vec u_2 + \vec u_3|$.  All the formal  relations
corresponding to  (\ref{7}),( \ref{8}), and (\ref{10})  are valid for four particles
as is expected.

\section{Nonlinear Equations for the Translationally Invariant Shell Model States}

 For three particles the simplest ansatz for a symmetrical space part wave function  is
 \begin{eqnarray}
 \Phi( u_1,u_2,u_3) = R(u_1) R(u_2) R(u_3),
\label{12}
 \end{eqnarray}
 with $ u_3 = | \vec u_1 + \vec u_2| $.

   It is straightforward, though tedious, to evaluate the action of the kinetic
energy in (\ref{9}) onto (\ref{12}). If we put $ R(u) = \frac{r (u)}{u} $,
    the Schr\"odinger equation  based on $ H_3 $ and the ansatz (\ref{12})
 results in
\begin{eqnarray}
   & & - \frac{1}{3m}[ r^{''}(u_1 ) r(u_2) r(u_3) + r(u_1) r^{''}(u_2) r(
 u_3) + r(u_1) r(u_2) r^{''}(u_3)\cr
   & - &  ( r^{'}( u_1) - \frac{ r(u_1)}{ u_1}) r(u_2) (  r^{'}( u_3) -
\frac{ r(u_3)}{ u_3})\hat u_1 \cdot \hat u_3\cr
    & - &   r( u_1) ( r^{'} (u_2)  - \frac{ r(u_2)}{ u_2})  (  r^{'}( u_3
) - \frac{ r(u_3)}{ u_3})\hat u_2 \cdot \hat u_3\cr
    & - &  ( r^{'}( u_1) - \frac{ r(u_1)}{ u_1})  (  r^{'}( u_2) - \frac{
 r(u_2)}{ u_2}) r(u_3) \hat u_1 \cdot \hat u_2]\cr
   &  + &   (V(u_1) + V(u_2) + V(u_3) - E) r(u_1) r(u_2 r(u_3) =0,
\label{13}
    \end{eqnarray}
where
\begin{eqnarray}
\hat u_1 \cdot \hat u_3 & = &  - \frac{ u_1 + \hat u_1 \cdot \vec u_2}{ u_3}\cr
\hat u_2 \cdot \hat u_3 & = &  - \frac{ u_2 + \hat u_2 \cdot \vec u_1}{ u_3}
 \end{eqnarray}
Here the independent variables are $ u_1,u_2 $ and $ x = \hat u_1 \cdot \hat u_2 $.

    We can not exclude that higher partial waves should be included.
  The  simplest ansatz for a  p-wave admixture is given by
   \begin{eqnarray}
   & & \Phi_1( \vec u_1,\vec u_2,\vec u_3)\cr
   &  = &  R_1( u_1) R_1( u_2) R(u_3) \hat u_1 \cdot \hat u_2
   +  R( u_1) R_1( u_2) R_1(u_3) \hat u_2 \cdot \hat u_3\cr
   &  + &   R_1( u_1) R( u_2) R_1(u_3) \hat u_1 \cdot \hat u_3,
\label{15}
    \end{eqnarray}
where for the sake of simplicity we assumed that the `third'  state, which is not
involved in the p-wave admixture, remains unchanged.  We leave it
to the reader to derive the resulting equation.

  For four particles the most simple ansatz is
  \begin{eqnarray}
  \Phi( u_1,u_2,u_3,u_4) = R(u_1)  R(u_2) R(u_3) R(u_4),
\label{16}
   \end{eqnarray}
   with $ | u_4 =| \vec u_1 + \vec u_2 + \vec u_3|$.
     The resulting equation based on $ H_4 $  is
\begin{eqnarray}
&  - &  \frac{3}{ 8m} [ r^{''}(u_1) r(u_2) r(u_3)   r(u_4)\cr
& + &  r (u_1) r (u_2) r (u_3)  r^{''}(u_4)  +  r(u_1)  r^{''}(u_2) r (u_
3)   r (u_4) \cr
& + & r (u_1) r (u_2) r (u_3)   r^{''}(u_4)  +  r (u_1) r ( u_2)  r^{''}(
u_3) r (u_4) \cr
& + & r (u_1) r ( u_2)  r (u_3)   r^{''}(u_4)\cr
&   - &    2 ( r'(u_1) - \frac{r(u_1)}{u_1}) r(u_2) r (u_3) ( r'(u_4) 
- \frac{r(u_4)}{u_4}) \hat u_1 \cdot \hat u_4\cr
& - &  2 r (u_1) ( r'(u_2) - \frac{r(u_2)}{u_2})  r (u_3) ( r'(u_4) 
- \frac{r(u_4)}{u_4}) \hat u_2 \cdot \hat u_4\cr
& - &  2 r (u_1)  r ( u_2) ( r'(u_3) - \frac{r(u_3)}{u_3})( r'(u_4)
 - \frac{r(u_4)}{u_4}) \hat u_3 \cdot \hat u_4 ]\cr
& - &  \frac{2}{3m} [  3  r (u_1) r (u_2) r (u_3) r^{''}(u_4) \cr
& + & ( r'(u_1) - \frac{r(u_1)}{u_1})( r'(u_2) - \frac{r(u_2)}{u_2})  r (u_3) r (u_4) 
\hat u_1 \cdot \hat u_2\cr
& - & ((  r (u_1) ( r'(u_2) - \frac{r(u_2)}{u_2})  r (u_3) ( r'(u_4) - \frac{r(u_4)}{u_4})\cr
&  + &   r (u_1) r (u_2) ( r'(u_3) - \frac{r(u_3)}{u_3}) ( r'(u_4) - \frac{r(u_4)}{u_4})\cr
&  + &   r (u_1) ( r'(u_2) - \frac{r(u_2)}{u_2}) r (u_3) ( r'(u_4) 
- \frac{r(u_4)}{u_4}))) \hat u_2 \cdot \hat u_4\cr
&  - & (( ( r'(u_1) - \frac{r(u_1)}{u_1}) r (u_2)  r (u_3) ( r'(u_4) - \frac{r(u_4)}{u_4})\cr
&  +  &  ( r'(u_1) - \frac{r(u_1)}{u_1}) r (u_2)  r (u_3) ( r'(u_4) 
- \frac{r(u_4)}{u_4}))) \hat u_1 \cdot \hat u_4\cr
& + & ( r'(u_1) - \frac{r(u_1)}{u_1}) r (u_2) ( r'(u_3)
 - \frac{r(u_3)}{u_3}) r (u_4) \hat u_1 \cdot \hat u_3\cr
&  - &  r (u_1)  r (u_2)  ( r'(u_3) - \frac{r(u_3)}{u_3}) ( r'(u_4) 
- \frac{r(u_4)}{u_4}) \hat u_3 \cdot\hat u_4\cr
& + & r (u_1) ( r'(u_2) - \frac{r(u_2)}{u_2})( r'(u_3) 
- \frac{r(u_3)}{u_3}) r (u_4) \hat u_2 \cdot \hat u_3\cr
& + & (  V(u_1) +  V(u_2)+   V(u_3)+   V(u_4))   r (u_1) r ( u_2) r (u_3) r (u_4)\cr
&  = &  E  r (u_1) r ( u_2) r (u_3) r (u_4).
\label{17}
    \end{eqnarray}

 Again, extensions to the ansatz (\ref{16}) are obvious. Already the presence of the
explicit angular dependence in (\ref{13}) and (\ref{17}) suggest that higher orbital
angular momentum admixtures are likely and that the most simple  ansatz for the
ground state may be poor.

 Choosing the mean field potential  $ V(u_i)$ to be a harmonic oscillator,
$ V(u_i) = \frac{ m \omega}{2} u_i^2 $, the nonlinear equations,  (\ref{13}) and
(\ref{17}), can be solved analytically.
  As  example we consider four  particles and introduce standard Jacobi coordinates
  \begin{eqnarray}
        \vec x &  = &  \vec x_2 - \vec x_3\cr
        \vec y & = &   \vec x_1 - \frac{1}{2} ( \vec x_2 + \vec x_3)\cr
        \vec z & = &  \vec x_4 - \frac{1}{3} ( \vec x_1 + \vec x_2 + \vec x_3)\cr
        \vec X & = &  \frac{1}{4} (  \vec x_1 + \vec x_2 + \vec x_3 + \vec x_4 ).
           \end{eqnarray}
           Then the potential energy $  V  =   \frac{m \omega}{2} (\frac{1}{2} x^2  + \frac{2}{3} y^2  + \frac{3}{4} z^2) $ as well as the kinetic energy
\begin{eqnarray}
        T_{rel}   =   -  \frac{1}{2m}( 2 \nabla_x^2 + \frac{3}{2} \nabla_y^2 + \frac{4}{3} \nabla_z^2 )
         \end{eqnarray}
  allow for a separation of the variables with the result
  \begin{eqnarray}
         \Phi( x,y,z) = e^{ - \frac{m \omega}{2}(  \frac{1}{2} x^2 + \frac{2}{3} y^2 + \frac{3}{4} z^2 )} = \prod_{i=1}^4 R(u_i).
\label{20}
\end{eqnarray}
           where
\begin{eqnarray}
R(u) = e^{ -  \frac{m \omega}{2} u^2}.
\end{eqnarray}
This goes with the lowest energy $ E= \frac{9}{2} \omega $.
The corresponding result for three particles, now for $ E= 3
 \omega $, is
\begin{eqnarray}
\Phi( x,y) =  \prod_{i=1}^3 R(u_i),
\label{22}
\end{eqnarray}
           with the same function $ R(u)$.
It is straightforward to verify that (\ref{20}) and (\ref{22})
fulfill the nonlinear equations (\ref{17}) and (\ref{13}).

\section{ The Faddeev-Yakubovsky Equations for Three and Four Particles}

The shell model Hamiltonians $ H_3 $  and $ H_4 $, Eqs.~(\ref{9}) and (\ref{11}),
 can be rewritten in terms of standard Jacobi coordinates.
This allows one to solve the two Schr\"odinger equations exactly in the form of the
Faddeev-Yakubovsky equations and therefore to test the quality of shell model ansatz.
          For three particles  one defines the Jacobi coordinates as
   \begin{eqnarray}
   \vec x & = &  \vec u_2 - \vec u_3 = 2 \vec u_2 + \vec u_1\cr
   \vec y & = &  \vec u_1 - \frac{1}{2} ( \vec u_2 + \vec u_3) = \frac{3}{2} \vec u_1,
\label{23}
      \end{eqnarray}
      or
\begin{eqnarray}
\vec u_1 & = &  \frac{2}{3} \vec y\cr
\vec u_2 & = &  \frac{1}{2} \vec x - \frac{1}{3} \vec y.
\label{23.2}
\end{eqnarray}
This gives for the Hamiltonian
 \begin{eqnarray}
 H_3 = - \frac{1}{m} \nabla_x^2 - \frac{3}{4m} \nabla_y^2 + V( \frac{ 2}{3} y) + V( | \frac{1}{2} \vec x - \frac{1}{3} \vec y|)
  + V( | \frac{1}{2} \vec x + \frac{1}{3} \vec y|)
 \end{eqnarray}
 The above expression has a formal similarity to a three-body Hamiltonian composed of
two-body  forces:
 \begin{eqnarray}
 H_{3,2b} = - \frac{1}{m} \nabla_x^2 - \frac{3}{4m} \nabla_y^2 + V_{2b}
( x) + V( | \frac{1}{2} \vec x +  \vec y|) + V( | \frac{1}{2} \vec x - 
\vec y|)  \, .
\label{23.22}
 \end{eqnarray}
However, Eqs. (\ref{23.2}) and (\ref{23.22}) are different.
   Nevertheless the formal structure  of the Faddeev equation \cite{fadd}
 can be used.
 The three-body bound state obeys
 \begin{eqnarray}
 \Psi = G_0 \sum_{i=1}^3 V_i \Psi \equiv \sum_{i=1}^3 \psi_i,
 \end{eqnarray}
 where $ G_0 $ represents the free three-body propagator, and $ V_i \equiv V( u_i)
 $. Then one arrives in a standard manner \cite{mybook} at
 \begin{eqnarray}
 \psi_i = G_0 T_i \sum_{j\ne i} \psi_j,
 \end{eqnarray}
 where $ T_i $ obeys the Lippmann  Schwinger equation
 \begin{eqnarray}
 T_i = V_i + V_i G_0 T_i\label{27}
 \end{eqnarray}
 Because of the identity of the particles one arrives at the well known form for
 the total state
 \begin{eqnarray}
 \Psi = ( 1 + P) \psi_1,
\label{28}
  \end{eqnarray}
  with $ P \equiv P_{12}P_{ 23} + P_{13}P_{23} $, which is a sum of a cyclical and
an anticyclical permutation of three particles.
  One Faddeev equation is sufficient, namely
  \begin{eqnarray}
  \psi_1 = G_0 T_1 P \psi_1\label{29}
  \end{eqnarray}

  The Faddeev equation can be solved in configuration space as an integro-differential
equation or, what we prefer, in momentum space as an integral equation. In the latter case
 one needs the momentum space representation  of the shell model potential as well as of
 the Lippmann Schwinger equation in terms of the conjugate momenta $\vec p_x $
and $ \vec p_y $ of the Jacobi momenta  $ \vec x $ and $ \vec y $.

With standard (unit) normalizations it results in
  \begin{eqnarray}
  < \vec x \vec y| \vec u_1 \vec u_2> = (\frac{1}{3})^3  \delta( \vec u
_1 - \frac{2}{3} \vec y) \delta( \vec u_2 - \frac{1}{2} \vec x + \frac{1}
{3} \vec y)\label{31}
  \end{eqnarray}

 Furthermore, as   consequence of the locality assumption
   \begin{eqnarray}
   < \vec u_1' \vec u_2'| V(u_1)| \vec u_1 \vec u_2> = \delta( \vec u_2
 - \vec u_2') \delta( \vec u_1 - \vec u_1') V(u_1)
    \end{eqnarray}
    and using (\ref{31}) one obtains
\begin{eqnarray}
< \vec p_x' \vec p_y'| V(u_1) | \vec p_x \vec p_y> = \delta( \vec
p_x - \vec p_x') \frac{ 1} { ( 2\pi)^3} \int d^3 y e^{ i( \vec p_y - \vec
 p_y') \cdot \vec y} V( \frac{2}{3} y).
\label{33}
\end{eqnarray}

       Due to that structure the T-matrix element in (\ref{27}) must have
 the form
\begin{eqnarray}
< \vec p_x' \vec p_y'| T_1 | \vec p_x \vec p_y> = \delta( \vec
p_x' - \vec p_x) t_1( \vec p_y', \vec p_y, z = E - \frac{ p_x^2}{m}),
\label{35}
\end{eqnarray}
         where $ t_1 $ obeys
\begin{eqnarray}
t_1 ( \vec p_y', \vec p_y, z) = V_1 ( \vec p_y', \vec p_y) + \int d^3 p_y^{''} V( \vec p_y', \vec p_y^{''}) 
\frac{1} { E - \frac{ p_x^2}{m} - \frac{3}{4m} p_y^{''2}} t_1( \vec p_y^{''}, \vec p_y, z).
\label{36}
\end{eqnarray}

         For two-body forces the $\delta$-function in (\ref{35}) would
have been for the spectator momentum $ \vec p_y $.
We assume that the mean field forces are spin-independent and require symmetry in the
spatial part.

  In \cite{liu1} such a system has been shown to be easily solvable using directly
momentum vectors and thus avoiding any partial wave decomposition.
We follow the same approach. Then (\ref{29}), using  (\ref{35}) has the form
  \begin{eqnarray}
 & &  < \vec p_x \vec p_y| \psi_1>  =   \frac{ 1}{ E - \frac{ p_x^2}{m}
 - \frac{3}{4m} p_y^2} \int d^3 p_{y'} t_1( \vec p_y, \vec p_y', z = E - \frac{ p_x^2}{m} )\cr
   & & \int d^3 p_{x^{''}} d^3 p_{ y^{''}} < \vec p_x  \vec p_{y'} | P | \vec p_{x^{''}} \vec p_{ y^{''}}>
    < \vec p_{x^{''}} \vec p_{ y^{''}}| \psi_1>.
\label{40}
\end{eqnarray}

     The permutation matrix element is well known \cite{mybook} and is given as
     \begin{eqnarray}
     < \vec p_x  \vec p_{y'} | P | \vec p_{x^{''}} \vec p_{ y^{''}}> & =
 &  (\frac{8}{3})^3 ( \delta( \vec p_{y'} + \frac{2}{3} \vec p_x + \frac{4}{ 3} \vec p_{ x^{''}})
     \delta( \vec p_{y^{''}} - \frac{4}{3} \vec p_x - \frac{2}{ 3} \vec p_{ x^{''}})\cr
     &  + &  \delta( \vec p_{y'} - \frac{2}{3} \vec p_x - \frac{4}{ 3} \vec p_{ x^{''}}) \delta( \vec p_{y^{''}}
      + \frac{4}{3} \vec p_x + \frac{2}{ 3} \vec p_{ x^{''}}))\label{38}
 \end{eqnarray}

 Therefore, Eq.~(\ref{40}) turns into
\begin{eqnarray}
  & & < \vec p_x \vec p_y| \psi_1>  =   \frac{ 1}{ E - \frac{ p_x^2}{m}
 - \frac{3}{4m} p_y^2} (\frac{8}{3})^3 \int d^3 p_{x{''}}\cr
  & & (  t_1( \vec p_y, - \frac{3}{2} \vec p_x - \frac{4}{3} \vec p_{x^{''}},
    z = E - \frac{ p_x^2}{m} ) < \vec p_{x^{''}}, \frac{4}{3} \vec p_x + \frac{2}{3} \vec p_{x^{''}}| \psi_1>  \cr
   &+  & t_1( \vec p_y,  \frac{3}{2} \vec p_x + \frac{4}{3} \vec p_{x^{''}},  z = E - \frac{ p_x^2}{m} ) < \vec p_{x^{''}}, - \frac{4}{3} \vec p_x - \frac{2}{3} \vec p_{x^{''}}| \psi_1>.
\label{42}
\end{eqnarray}

     Because of the uniqueness of the solution, any solution of  (\ref{42}) has
the property $ < - \vec p_x, \vec p_y| \psi_1> = <  \vec p_x, \vec p_y| \psi_1> $.
This equation is can then be solved by iteration using a Lanczos type algorithm \cite{stadler}.

As follows from (\ref{38}) the total state  given by (\ref{28})  has the
 form
\begin{eqnarray}
< \vec p_x, \vec p_y| \Psi>  & = &  < \vec p_x, \vec p_y| \psi_1> + ( \frac{3}{4})^3 
 < - \frac{1}{2} \vec p_x -  \frac{3}{4} \vec p_y, \vec p_x
 +  \frac{1}{2} \vec p_y | \psi_1>\cr
& + & ( \frac{3}{4})^3  < - \frac{1}{2} \vec p_x +  \frac{3}{4} \vec p_y,
 - \vec p_x +  \frac{1}{2} \vec p_y | \psi_1>.
\label{49}
 \end{eqnarray}

In the case of four particles we use the Yakubovsky equations \cite{yakub}. For
four bosons and two-body forces this has been solved rigorously the first time in \cite{kamada2}.
Now we have different  potentials depending on the relative coordinates $\vec u_i$,
which require a renewed derivation. Starting from
 \begin{eqnarray}
 \Psi = G_0 \sum_{ i=1}^4 V( u_i) \Psi \equiv \sum_{ i=1}^4 \psi_i
 \end{eqnarray}
  one arrives in a standard first step at
  \begin{eqnarray}
  \psi_1 = G_0 T_1 ( \psi_2+ \psi_3+ \psi_4),
  \end{eqnarray}
  where $ T_1 $ obeys the Lippmann Schwinger equation (\ref{36}).
(Note however, the modified free four-body propagator.)

  In the spirit of the Yakubovsky scheme one  regards a three- body subsystem by defining
   \begin{eqnarray}
   \psi_{1; 123} \equiv G_0 T_1 ( \psi_2 + \psi_3)\label{43}
    \end{eqnarray}
    and a remaining component
   \begin{eqnarray}
   \psi_{1; 1,4} \equiv G_0 T_1  \psi_4 .
\label{44}
    \end{eqnarray}
    Then
\begin{eqnarray}
\psi_1 = \psi_{1; 123} +  \psi_{1; 1,4}.
\end{eqnarray}
      Correspondingly one defines
   \begin{eqnarray}
   \psi_{2; 231} & = &  G_0 T_2 ( \psi_3 + \psi_1)\cr
      \psi_{3; 312} & = &  G_0 T_3 ( \psi_1 + \psi_2)\cr
    \psi_{2; 2,4} & = &  G_0 T_2 \psi_4\cr
     \psi_{3; 3,4} & = &  G_0 T_3 \psi_4
   \end{eqnarray}
    where
    \begin{eqnarray}
   \psi_2 & = &  \psi_{2; 231} +  \psi_{2; 2,4}\cr
   \psi_3 & = &  \psi_{3; 312} +  \psi_{3; 3,4}.
\label{47}
    \end{eqnarray}
      Then (\ref{43}) and ( \ref{47}) yield
   \begin{eqnarray}
   \psi_{ 1; 123} = G_0 T_1 ( \psi_{ 2; 231} + \psi_{ 3; 312} + \psi_{
2; 2,4} + \psi_{3; 3,4}).
\label{48}
    \end{eqnarray}
     Due to the identity of the particles   one has
    \begin{eqnarray}
   \psi_{ 2; 231} +  \psi_{ 3; 312} & = &  P \psi_{1;123}\cr
   \psi_{ 3; 3,4} & = &  P_{23} \psi_{ 2; 2,4},
      \end{eqnarray}
      and (\ref{48}) can be rewritten as
\begin{eqnarray}
( 1 - G_0 T_1 P) \psi_{1;123} = G_0 T_1( 1 +  P_{23})  \psi_{ 2; 2 ,4}.
\end{eqnarray}
      The left hand side by itself defines a three-body problem. After inversion one obtains
\begin{eqnarray}
\psi_{1;123} = G_0 \hat{T} ( 1 +  P_{23})  \psi_{ 2; 2,4},
\end{eqnarray}
where $ \hat{T} $ obeys
\begin{eqnarray}
\hat {T} = T_1 + T_1 P G_0  \hat {T}.
\end{eqnarray}

It remains to consider  (\ref{44}), which in analogy to (\ref{47}) has the form
\begin{eqnarray}
\psi_{2; 2,4} = G_0 T_2 ( \psi_{ 4; 413} + \psi_{ 4; 4,2}).
\label{56}
\end{eqnarray}
Using now
\begin{eqnarray}
\psi_{ 4; 4,2} = P_{ 24} \psi_{ 2; 2,4}
\end{eqnarray}
we rewrite (\ref{56}) as
\begin{eqnarray}
( 1 - G_0 T_2 P_{24})   \psi_{ 2; 2,4} = G_0 T_2 \psi_{4; 413}.
\end{eqnarray}
Inversion yields
\begin{eqnarray}
\psi_{ 2; 2,4} = G_0  \tilde {T} \psi_{4; 413},
\end{eqnarray}
where $ \tilde { T} $ obeys
 \begin{eqnarray}
  \tilde {T} = T_2 + T_2 P_{24} G_0  \tilde { T}
   \end{eqnarray}

   Finally permutation symmetry yields
   \begin{eqnarray}
   \psi_{4;423} = P_{23}P_{14} \psi_{1; 123}
    \end{eqnarray}
    and one ends up with two coupled equations
    \begin{eqnarray}
 \psi_{1;123} & = &  G_0 \hat {T} ( 1 + P_{23}) \psi_{2; 2,4}\cr
 \psi_{2; 2,4} & = &  G_0 \tilde {T}  P_{23}P_{ 14} \psi_{1;123}.
\label{59}
 \end{eqnarray}

    The total wave function is now given as
     \begin{eqnarray}
     \Psi & = &  \psi_1 + \psi_2 + \psi_3 + \psi_4  =   ( 1 + P) \psi
_{1;123} +  P_{14} P_{23}( \psi_{1;123}\cr
&  + &  \psi_{2;2,4}) + ( 1 + P_{24} + P_{12}) \psi_{ 2;2,4}.
\label{60}
 \end{eqnarray}

     While a corresponding coupled set based on two- body forces has been rigorously
solved~\cite{kamada2} in a partial wave representation, it is also possible to directly use
momentum vectors as has been demonstrated in \cite{teheran}.

     We would propose to follow that second option.
We leave it to the reader to work out the explicit momentum space representation of
(\ref{59}) and (\ref{60}) in terms of appropriate Jacobi momentum vectors.

\section{Shell Model Ansatz versus Exact Wave Function}

        The solution of the Faddeev equation (\ref{42}) yields the full
three- dimensional three-boson Faddeev component in momentum space. This
is the input for the full wave function
         given in (\ref{49}). Since we search for the lowest energy state,
$ \Psi$  is a scalar and therefore depends only on 3 variables
\begin{eqnarray}
< \vec p_x, \vec p_y| \Psi> \rightarrow \Psi( p_x, p_y, \hat p_x \cdot \hat p_y)
\end{eqnarray}

         As a consequence, the dependence of the configuration space wave
function \mbox{ $ < \vec x \vec y| \Psi> $} will also reduce to a three-variable dependence
$\Psi( x, y,\hat x \cdot \hat y)$:
\begin{eqnarray}
< \vec x \vec y| \Psi> & = &  \frac{1}{(2\pi)^3} \int d^3 p_x
d^3 p_y e^{i( \vec p_x \cdot \vec x + \vec p_y \cdot \vec y)} \Psi( p_x,
p_y,\hat p_x \cdot \hat p_y)\cr
& = &  \frac{1}{(2\pi)^3} \int d^3 p_x d^3 p_y cos ( \vec p_x
 \cdot \vec x + \vec p_y \cdot \vec y) \Psi( p_x, p_y,\hat p_x \cdot \hat
 p_y)\cr
& & \equiv \Psi( x,y, \hat x \cdot  \hat y).
\end{eqnarray}

We used the reality property of $\Psi$ to replace the exponential by the cosine.

The expectation is now that
\begin{eqnarray}
\Psi_{SM} ( \vec x,\vec y) \equiv  R(u_1)  R(u_2) R(u_3),
\end{eqnarray}
with $ u_3 = | \vec u_1 + \vec u_2| $ being a good approximation to
$ \Psi( x, y, \hat x \cdot \hat y) $. In the case of the harmonic oscillator this
is exactly fulfilled.

  In general one faces the task to minimize
   $|\Psi( x, y, \hat x \cdot \hat y) -  R(u_1)  R(u_2) R(u_3)| $ for all $ x, y,  \hat x \cdot \hat y $
or $ u_1, u_2,  \hat u_1 \cdot \hat u_2 $. Explicitly this requirement  is
  $ | \Psi( x, y,\hat x \cdot \hat y) - R(\frac{2}{3} y) R( | \frac{1}{2} \vec x 
  - \frac{1}{3} \vec y|) R( | \frac{1}{2} \vec x + \frac{1}{3} \vec y|)| $ or
  $ | \Psi( | \vec u_1 + 2 \vec u_2|, \frac{3}{2} u_1, \frac{ u_1 + 2 \vec u_2 \cdot \hat u_1} {|  \vec u_1 + 2 \vec u_2|} - R(u_1) R(u_2) R(| \vec u_1 + \vec u_2|) $ to be minimal.

   Instead of an optimized pointwise adjustment one can try an average adjustment minimizing
\begin{eqnarray}
        & & \int du_1 du_2 d \hat u_1 \cdot \hat u_2 ( \Psi( | \vec u_1 +
 2 \vec u_2|, \frac{3}{2} u_1, \frac{ u_1 + 2 \vec u_2 \cdot \hat u_1} {|
  \vec u_1 + 2 \vec u_2|})\cr
         & - &  R(u_1) R(u_2) R(| \vec u_1 + \vec u_2|))^2
\end{eqnarray}
          in relation to the choice of $ R(u) $. For instance, one can expand
$R(u) $ into harmonic oscillator wave functions $ \Phi_m(u) $, where
 $\frac{m \omega}{2} $ is optimally adjusted to the given mean field potential $V(u)$.

        Thus
\begin{eqnarray}
R(u) = \sum_m \phi_m (u) C_m,
\end{eqnarray}
and the set ${ C_m} $ is to be varied minimizing the above integral.
Differentiating  with respect to $ C_k $ and putting the result
to zero yields a nonlinear relation for the coefficients $ C_m $.
This might be solved by an iterative procedure allowing first
 $ C_0 \ne 0 $. Then keeping also $ C_1 \ne 0 $ in addition one might start
with $ C_0 $ from the previous step and determine $ C_1 $. Finally
one can iterate the nonlinear equation for $ C_0 $ and $ C_1$  starting with
the values found before; etc.
Very likely, however, one has to allow in addition for p-wave admixtures
as given in (\ref{15}) and possibly even higher orbital  angular momentum values.

The direct solution of the nonlinear equation (\ref{13}) poses a
severe problem. Moreover, very likely p-wave and possibly higher order
admixtures have to be taken  into account, which requires an extension of
 the nonlinear equation (\ref{13}) as mentioned above. Discretization in
the $ u_1,u_2, \hat x \cdot \hat y $ - values is necessary and iterative
procedures appear unavoidable. Thereby each run is of course an eigenvalue
problem for the energy $ E $.

In the case of four nucleons the symmetric state of lowest energy is again a scalar
and thus depends on 5 variables:
\begin{eqnarray}
\Psi = \Psi( x, y, z, \hat x \cdot \hat y, \hat x \cdot \hat z, \hat y
\cdot \hat z),
\end{eqnarray}
  where $ \vec x, \vec y, \vec z $ are one choice of standard Jacobi coordinates.
The optimal extraction of $ R(u) $ in
\begin{eqnarray}
\Psi_{SM} ( \vec x, \vec y, \vec z) \equiv R( u_1)  R( u_2) R( u_3) R( u_
4)
\end{eqnarray}
and possibly higher angular momentum admixture
  follows analogous strategies as for three nucleons.

\section{ Realistic Three- and Four-Nucleon Wave Functions}

  Based on modern nuclear forces like~\cite{av18,cdbonn,nijm} combined
with three-nucleon (3N) forces of the Tucson-Melbourne type~\cite{TM3NF} or based on
   the most recent consistent  two- and three-nucleon forces  generated from
chiral effective field theory~\cite{epel} numerically exact solutions of
the Faddeev - Yakubovsky equations are available. If a correlated single particle
picture applies at all it can only be valid beyond a certain value $r_0$  of the
   pair distances. The two-body correlation  function to find two nucleons at a distance  $r$
has its maximum around $r=1$~fm  universally for all light nuclei~\cite{gloeck2,gloeck3,gloeck4}.
    Thus $r_0$ has to be smaller than  1~fm. For the most simple
correlated shell model ansatz of Eq.~(\ref{12}) or symmetric extensions
beyond s-wave and (\ref{16}) the exact wave function for $^3$He  and $^4$He
is to be projected onto the totally  antisymmetric spin-isospin states
$\chi_3 $ and $\chi_4 $, Eqs.~(\ref{5}) and (\ref{6}),  respectively:
\begin{eqnarray}
\Psi_{3,4}^{exact} \equiv < \chi_{3,4}| \Psi_{3,4}^{exact}>.
\end{eqnarray}
     For a global adjustment one has to minimize
 \begin{eqnarray}
 \int dV (\Psi_{3,4}^{exact} - \prod_{i=1}^{3,4} R(u_i) )^2 \prod_{i<j}
^{3,4} \Theta ( r_{ij} - r_0)
 \end{eqnarray}
 or an extension including higher partial waves but still keep the symmetry in the space part.

 The resulting $ R(u) $  and $ R's $ related to higher partial waves should
be independent of $ r_0 $. This requirement should determine the smallest possible value
for $r_0$.

 Knowing $ R(u) $ one can compare the norms
 \begin{eqnarray}
 N^{exact} & \equiv &  \int dV |  \Psi_{3,4}^{exact}|^2 \prod_{i<j}^{3,4}
 \Theta ( r_{ij} - r_0)\cr
 N^{SM} & \equiv &   \int dV  | \prod_{i=1}^{3,4} R(u_i) |^2  \prod_{i<
j}^{3,4} \Theta ( r_{ij} - r_0)
 \end{eqnarray}

   In addition the  short range behavior is not accessible to the single
particle picture and provides the norm contribution
    \begin{eqnarray}
 N_{short}^{exact}  \equiv   \int dV |  \Psi_{3,4}^{exact}|^2 \prod_{i<j}
^{3,4} \Theta ( r_0 - r_{ij})
 \end{eqnarray}
Finally, one has to keep in mind that only about 90\%
 of the total norm is related to the spin-isospin states $ \chi_{3,4} $.
The rest is of more complicated structure~\cite{nogga5}.

\section{Summary}

  In nature a nuclear wave function is translationally invariant. Therefore,
 if a shell model picture is a good representation of a nuclear wave function, the single
   particle states have to depend on translationally invariant coordinates.
Our choice of coordinates $\vec u_i \equiv \vec x_i - \vec X $ relating the individual position
    vectors $ \vec x_i $ to the c. m. coordinate $ \vec X $ fulfills this
condition with the additional constraint that they have to sum up to zero:
  $ \sum_{i=1}^n \vec u_i =0 $.
     Choosing the first $ n - 1$ of them together with the c. m. coordinate one
can formulate a shell model Hamiltonian composed of kinetic energy
 containing now also mixed
      terms $ \vec \nabla_{u_i} \cdot \vec \nabla_{u_j} $ and single  particle potentials
depending on the  coordinates $ | u_i| $. Assuming the
energetically lowest energy state
       to that Hamiltonian to be a Slater determinant with equal space dependent
single particle wave functions, $ R(u_i) $, which is the most simple choice,
one obtains nonlinear equations for $ R(u_i) $. They have been worked out
for nucleon numbers $A=3$  and 4. For the special choice
of harmonic oscillator potentials  the nonlinear equations can be analytically
solved and that most simple ansatz for the wave function turns
         out to be correct. In the case of general mean field potentials  partial
wave contributions beyond s-states might be necessary.

    In order to shed light on the  question how well such a shell model
ansatz is justified we regarded in some detail three and four nucleons. The
corresponding shell model Hamiltonian can be written in terms of standard
Jacobi coordinates and numerically exact solutions can be generated based on
the  Faddeev-Yakubovsky equations.  Knowing the exact wave functions
one can check the validity  of the Slater determinant ansatz.   Optimization
algorithms are provided to perform the  comparison of exact wave function
with the Slater determinant ansatz.

      The main task however,  is to confront such a shell model ansatz to
realistic three-  and four-nucleon wave functions (and beyond), which are based on
modern  two- and three-nucleon forces.
      Clearly at short pair distances the well established repulsive nature of
the  nuclear forces invalidates the shell model ansatz and therefore only
for pair distances beyond a certain value $ r_0 $ the shell  model picture can
make sense, if at all.   To that aim numerical investigations are planned for both,
the shell-model Hamiltonians and realistic Hamiltonians
composed of two- and three-nucleon forces.

\end{document}